\documentclass[longauth]{aaemma} 
\usepackage{graphicx,color}
\usepackage{natbib}
\bibpunct{(}{)}{;}{a}{}{,} 

\usepackage[small]{caption}

\begin{document}
%

\newcommand\arcdeg{$^{\circ}$}
\newcommand\arcs{$^{\prime\prime}$}
\newcommand{\arcm}{$^\prime$}
\newcommand\wtwo{Westerlund~2}

\title{Revisiting the Westerlund~2 Field with the H.E.S.S. Telescope Array}

\subtitle{}

%
\author{ HESS Collaboration
 \and A.~Abramowski \inst{4}
 \and F.~Acero \inst{15}
 \and F. Aharonian\inst{1,13,35}
 \and A.G.~Akhperjanian \inst{2,35}
 \and G.~Anton \inst{16}
 \and A.~Barnacka \inst{24,7}
 \and U.~Barres de Almeida \inst{8} \thanks{supported by CAPES Foundation, Ministry of Education of Brazil}
 \and A.R.~Bazer-Bachi \inst{3}
 \and Y.~Becherini \inst{12}
 \and J.~Becker \inst{21}
% ...
\newpage
}
\institute{
Max-Planck-Institut f\"ur Kernphysik, P.O. Box 103980, D 69029
Heidelberg, Germany
\and
 Yerevan Physics Institute, 2 Alikhanian Brothers St., 375036 Yerevan,
Armenia
\and
Centre d'Etude Spatiale des Rayonnements, CNRS/UPS, 9 av. du Colonel Roche, BP
4346, F-31029 Toulouse Cedex 4, France
\and
Universit\"at Hamburg, Institut f\"ur Experimentalphysik, Luruper Chaussee
149, D 22761 Hamburg, Germany
\and
Institut f\"ur Physik, Humboldt-Universit\"at zu Berlin, Newtonstr. 15,
D 12489 Berlin, Germany
\and
LUTH, Observatoire de Paris, CNRS, Universit\'e Paris Diderot, 5 Place Jules Janssen, 92190 Meudon, 
France
\and
CEA Saclay, DSM/IRFU, F-91191 Gif-Sur-Yvette Cedex, France
\and
University of Durham, Department of Physics, South Road, Durham DH1 3LE,
U.K.
\and
Unit for Space Physics, North-West University, Potchefstroom 2520,
    South Africa
\and
Laboratoire Leprince-Ringuet, Ecole Polytechnique, CNRS/IN2P3,
 F-91128 Palaiseau, France
\and 
Laboratoire d'Annecy-le-Vieux de Physique des Particules,
Universit\'{e} de Savoie, CNRS/IN2P3, F-74941 Annecy-le-Vieux,
France
\and
Astroparticule et Cosmologie (APC), CNRS, Universit\'{e} Paris 7 Denis Diderot,
10, rue Alice Domon et L\'{e}onie Duquet, F-75205 Paris Cedex 13, France
\thanks{UMR 7164 (CNRS, Universit\'e Paris VII, CEA, Observatoire de Paris)}
\and
Dublin Institute for Advanced Studies, 5 Merrion Square, Dublin 2,
Ireland
\and
Landessternwarte, Universit\"at Heidelberg, K\"onigstuhl, D 69117 Heidelberg, Germany
\and
Laboratoire de Physique Th\'eorique et Astroparticules, 
Universit\'e Montpellier 2, CNRS/IN2P3, CC 70, Place Eug\`ene Bataillon, F-34095
Montpellier Cedex 5, France
\and
Universit\"at Erlangen-N\"urnberg, Physikalisches Institut, Erwin-Rommel-Str. 1,
D 91058 Erlangen, Germany
\and
Laboratoire d'Astrophysique de Grenoble, INSU/CNRS, Universit\'e Joseph Fourier, BP
53, F-38041 Grenoble Cedex 9, France 
\and
Institut f\"ur Astronomie und Astrophysik, Universit\"at T\"ubingen, 
Sand 1, D 72076 T\"ubingen, Germany
\and
LPNHE, Universit\'e Pierre et Marie Curie Paris 6, Universit\'e Denis Diderot
Paris 7, CNRS/IN2P3, 4 Place Jussieu, F-75252, Paris Cedex 5, France
\and
Charles University, Faculty of Mathematics and Physics, Institute of 
Particle and Nuclear Physics, V Hole\v{s}ovi\v{c}k\'{a}ch 2, 
180 00 Prague 8, Czech Republic
\and
Institut f\"ur Theoretische Physik, Lehrstuhl IV: Weltraum und
Astrophysik,
    Ruhr-Universit\"at Bochum, D 44780 Bochum, Germany
\and
University of Namibia, Department of Physics, Private Bag 13301, Windhoek, Namibia
\and
Obserwatorium Astronomiczne, Uniwersytet Jagiello{\'n}ski, ul. Orla 171,
30-244 Krak{\'o}w, Poland
\and
Nicolaus Copernicus Astronomical Center, ul. Bartycka 18, 00-716 Warsaw,
Poland
 \and
School of Physics \& Astronomy, University of Leeds, Leeds LS2 9JT, UK
 \and
School of Chemistry \& Physics,
 University of Adelaide, Adelaide 5005, Australia
 \and 
Toru{\'n} Centre for Astronomy, Nicolaus Copernicus University, ul.
Gagarina 11, 87-100 Toru{\'n}, Poland
\and
Instytut Fizyki J\c{a}drowej PAN, ul. Radzikowskiego 152, 31-342 Krak{\'o}w,
Poland
\and
Astronomical Observatory, The University of Warsaw, Al. Ujazdowskie
4, 00-478 Warsaw, Poland
\and
Institut f\"ur Astro- und Teilchenphysik, Leopold-Franzens-Universit\"at 
Innsbruck, A-6020 Innsbruck, Austria
\and
Oskar Klein Centre, Department of Physics, Stockholm University,
Albanova University Center, SE-10691 Stockholm, Sweden
\and
Oskar Klein Centre, Department of Physics, Royal Institute of Technology (KTH),
Albanova, SE-10691 Stockholm, Sweden
\and
Department of Physics and Astronomy, The University of Leicester, 
University Road, Leicester, LE1 7RH, United Kingdom
\and
European Associated Laboratory for Gamma-Ray Astronomy, jointly
supported by CNRS and MPG
\and
National Academy of Sciences of the Republic of Armenia, Yerevan 
% ...
}

\newpage

\offprints{\\emma@mpi-hd.mpg.de,\\Olaf.Reimer@uibk.ac.at}

\date{Received  ; accepted }

\abstract
{}
{Previous observations with the H.E.S.S. telescope array revealed the
  existence of extended very-high-energy (VHE; E$>$100~GeV) $\gamma$-ray emission, \object{HESS J1023--575}, coincident with the
  young stellar cluster Westerlund~2. 
At the time of discovery, the origin of the observed emission was not
unambiguously identified, and follow-up observations have been
performed to further investigate the nature of this $\gamma$-ray source.}
{The Carina region towards the open cluster Westerlund 2 has been
  re-observed, increasing the total exposure to 45.9~h. The combined dataset includes 33~h of new data and now permits a search for energy-dependent morphology and detailed spectroscopy.}
{A new, hard spectrum VHE $\gamma$-ray source, \object{HESS\,J1026--582}, was discovered with a statistical significance of 7$\sigma$. It is positionally coincident with the Fermi LAT pulsar \object{PSR\,J1028--5819}. The
    positional coincidence and radio/$\gamma$-ray characteristics of
    the LAT pulsar favors a scenario where the TeV emission originates from a pulsar wind nebula. The nature of HESS\,J1023--575 is discussed in light of the deep H.E.S.S. observations and recent multi-wavelength discoveries,
  including the Fermi LAT pulsar \object{PSR\,J1022--5746} and giant molecular
  clouds in the region. Despite the improved VHE dataset, a clear
  identification of the object responsible for the VHE emission from HESS\,J1023--575 is not yet possible, and contribution
  from the nearby high-energy pulsar and/or the open cluster remains a possibility.  
} 
{}

\authorrunning{H.E.S.S. Collaboration }
\titlerunning{}
\maketitle

\section{Introduction}

The H.E.S.S. (High Energy Stereoscopy System) collaboration reported the detection of a bright, extended
(standard deviation $\sigma$=0.18$^{\rm o}$) VHE $\gamma$-ray source HESS\,J1023--575 \citep{HESSJ1023m575}
coincident with the open stellar cluster Westerlund~2. This discovery
provides evidence for particle acceleration to extremely high energies
in this region. 
Young stellar clusters are among the sites considered for high-energy
$\gamma$-ray production. This relates to the existence of late states of
stellar evolution, which are generally considered favorable for
particle acceleration. Different scenarios have been proposed to
account for VHE $\gamma$-ray emission from stellar environments,
considering the contribution of hot and massive Wolf-Rayet (WR) and
OB stars and their winds in single, binary, or collective processes,
pulsars and their synchrotron nebulae, as well as supernova
explosions and their expanding remnants
\citep{lepwinds1, lepwinds2, lepwinds3, lepwinds4, PandD, torres06, Bed05, Horns}. The
giant \ion{H}{II} region \object{RCW~49} (NGC~3247) and
its ionizing cluster \object{Westerlund~2}, hosting over two dozen of massive
stars as well as two remarkable WR stars (WR~20a and WR~20b), have been studied
extensively over the whole electromagnetic spectrum, and certainly
interest was further boosted by the discovery of VHE $\gamma$-ray
emission from the vicinity of Westerlund 2.

The VHE $\gamma$-ray source HESS J1023--575 \citep{HESSJ1023m575} is characterized by a
power-law spectrum with a photon index $\Gamma \approx 2.5$ and, at a nominal distance of 8~kpc (obtained
from timing studies of the WR binary WR 20a; \citealt{Rauw07}), would
have a $\gamma$-ray luminosity $L_{\gamma}$(E $>$ 380~GeV) = 1.5$\times$10$^{\rm 35}$ erg s$^{-1}$. The extent of the observed VHE $\gamma$-ray
emission spatially overlaps with the wind-blown bubbles observed in
infrared (IR) by \emph{Spitzer} \citep{Spitzer} and in radio continuum
by ATCA at 1.38 and 2.38 GHz \citep{radiocontinuum}. One of these
bubbles, with a diameter of $\sim$7.3\arcm\, is believed to be connected
to the massive binary WR~20a, surrounding the central region of
Westerlund~2, while the second has been related to WR~20b, with a
diameter of $\sim$4.1\arcm. The WR binaries WR~20a and WR~20b have been detected in X-rays in the 0.5 to 10 keV band with \emph{Chandra} \citep{Chandra} with luminosities of 19$\times$10$^{\rm
  32}$~erg s$^{-1}$ and 52$\times$10$^{\rm 32}$~erg s$^{-1}$ respectively. The high X-ray luminosity seen
from WR~20a, which is likely associated with the open cluster,
provides support for the scenario whereby in a colliding wind zone energy dissipated in
shocks is converted into thermal energy \citep{Chandra}. The observed
source extension (0.18\arcdeg) and non-variability of the $\gamma$-ray flux of
HESS\,J1023--575 disfavor any scenario where particle acceleration
occurs close to the massive stars, and in the colliding wind zone
within the massive WR~20a binary system scenario.
A prominent feature at radio wavelengths, known as the "blister"
\citep{radiocontinuum}, coincident with HESS\,J1023--575 \citep{HESSJ1023m575}, does not seem to be evident in the diffuse
X-ray flux measured by \emph{Chandra} \citep{Chandra}, which appears
featureless and predominantly of thermal origin. Upper limits on the
non-thermal diffuse emission were recently reported by \cite{SUZAKU}
in the 0.7--2 keV (F$_{\rm ul}$$<$2.6$\times$10$^{\rm -12}$ erg cm$^{\rm -2}$ s$^{\rm -1}$) and 0.7 to 10 keV
(F$_{\rm ul}$$<$6.1$\times$10$^{\rm -12}$ erg cm$^{\rm -2}$ s$^{\rm
  -1}$) bands, using \emph{Suzaku} observations and integrating over the entire camera field (17.8\arcm$\times$17.8\arcm). According
to \cite{SUZAKU}, the statistics are not sufficient to
distinguish between three thermal components and two thermal components
plus a non-thermal one when modeling the background. They claim that their multicomponent fit supports the existence of non-thermal emission. However, the ambiguity in the goodness-of-fit
using a multi-thermal component and a multi-thermal component including a non-thermal one does not warrant
such a conclusion on the basis of the presented data.

A possible scenario discussed to explain the VHE $\gamma$-ray
emission includes the interaction of cosmic-ray protons accelerated in
stellar wind shocks or supernova blast waves with dense molecular
clouds \citep{AharonianAtoyan96}. The molecular cloud content in this
region observed by the CfA 1.2~m Millimeter-Wave Telescope and the 4-m
Nanten 2 sub-mm telescope shows a complex and rich environment. \cite{DameCO} observed Westerlund 2 with the CfA 1.2~m telescope using $^{\rm
  12}$CO(J~=~1--0) survey data and found a possible association with a
giant molecular cloud (GMC) on the far side of the Carina spiral arm, at a
distance of 5.0 kpc. Observations at a higher resolution (90\arcs) on the
J~=~2--1 transition of $^{\rm 12}$CO performed by the Nanten 2 4-m
sub-millimeter telescope \citep{MCNANTEN_2,MCNANTEN_1} confirm the \cite{DameCO}
results but also suggest another possible association. A significant
clump of molecular gas, consisting of a foreground component at 0~km
s$^{-1}$ is found, with a mass of (9.1$\pm$4.1)$\times$10$^4$ M$_{\odot}$, and a background component at
~16~km s$^{-1}$, with a mass of (8.1$\pm$3.7) $\times$10$^4$ M$_{\odot}$,
both of which are claimed to be connected with RCW~49, at a distance of
5.4$^{\rm +1.1}_{-1.4}$ kpc. The analysis of the Nanten survey in the
$^{\rm 12}$CO (J~=~1--0) \citep{Fukui} also shows two unusual
structures in the vicinity of the line-of-sight projection: a straight
feature in the East and an arc-like feature in the West in the
velocity range 20--30~km s$^{-1}$, corresponding to a distance of
$\sim$7~kpc (using the rotation curve of \citealt{BandB93}), although the
association with the open cluster is unclear.  

At high energies (HE; E$>$100~MeV) the Fermi Large Area Telescope
(LAT) collaboration has recently reported the detection of two new
sources: \object{1FGL~J1023.0--5746}, positionally coincident with
HESS\,J1023--575, and a new $\gamma$-ray pulsar \object{1FGL\,J1028.4--5810}
(PSR\,J1028--5819) to the east of the H.E.S.S. source and associated
with the EGRET source \object{3EG\,J1027--5817} \citep{fermi1028}. 

The first source, reported in the bright source list, has been very recently identified with a new
$\gamma$-ray pulsar, PSR\,J1022--5746 \citep{Fermisymp,fermi1cat,FermiBlind}. The
pulsar periodicity found in the blind search  performed by the Fermi
LAT collaboration is 111.47 ms. The young pulsar (with a characteristic age of
$\tau\sim$4.6 kyr) has a spin-down luminosity of 1.1$\times$10$^{37}$ erg
s$^{-1}$. Radio pulsations have not been found but the 130 ksec \emph{Chandra} 
image shows a faint source (CXOU J102302.8--574606.9, \citealt{chandrapsr}) identified as a possible
counterpart. 
The second source, PSR\,J1028--5819 \citep{fermi1028,ATNF} is a 91.4~ms $\gamma$-ray and radio pulsar
with a total spin-down power of $\dot{E}$ = 8.3$\times$10$^{\rm 35}$
erg~s$^{-1}$ and a phase-average luminosity from 0.1 to 30~GeV of (120$\pm$73)$\times$10$^{\rm 33}$ erg
s$^{\rm -1}$ at a dispersion measure-derived distance of 2.3$\pm$0.7 kpc. Its characteristic age
is 8.9$\times$10$^4$ yrs. The HE photon spectrum
shows a hard cutoff at 1.9$\pm$0.5 GeV, while no contribution of
steady emission is reported. 

The unresolved origin of the VHE $\gamma$-ray emission from the
direction of Westerlund~2 motivated
further observations, aiming to discriminate among the alternative
scenarios for the VHE emission. The results of these deep observations, alongside the discovery of a
new, hard, VHE $\gamma$-ray emission coincident with the Fermi LAT
source PSR\,J1028--5819, are reported here. The nature of the VHE
sources is then discussed in light of the new multi-wavelength
observations in the Westerlund~2 field.

%
\section{Analysis and Results}
\label{ANALYSIS}

The High Energy Stereoscopic System (H.E.S.S.) is an array of
four VHE $\gamma$-ray imaging atmospheric Cherenkov telescopes (IACTs) located in the Khomas Highland of Namibia (23$^{\rm o}$16\arcm18\arcs
S 16$^{\rm o}$30\arcm00\arcs\ E). Each of these telescopes is equipped with a
tessellated spherical mirror of 107~m$^2$ area and a camera comprised
of 960 photomultiplier tubes, covering a large field-of-view (FoV) of
5$^{\rm o}$ diameter. Its large FoV and good off-axis sensitivity make
H.E.S.S. ideally suited for studies of extended sources and regions of the
sky in which more than one $\gamma$-ray source could be present (see
e. g. \citealt{HESSJ1259}). The system works in a coincidence mode
(e.g. \citealt{trigger}), requiring at least two of the four telescopes
to trigger the detection of an extended air shower (EAS). This
stereoscopic approach results in a high angular
resolution of $\sim$~5\arcm\ per event, good energy resolution (6$\%$ on
average) and an effective background rejection \citep{HESSCrab}. These characteristics
allow H.E.S.S. to reach a sensitivity
$\sim$2.0$\times$10$^{-13}$~ph~cm$^{-2}$s$^{-1}$ (equivalent to 1$\%$
of the Crab Nebula flux above 1~TeV), or less if advanced techniques are used for image
analysis \citep{NGC253}, for a point-like source detected at a
significance of 5$\sigma$ in 25~hours of observation at zenith. 

Previous observations of the region containing PSR\,J1028--5819 and
Westerlund~2 \citep{HESSJ1023m575} were performed between March and
July 2006, amounting to a total of 12.9~h of live time. The
analysis of the data led to the discovery of an extended source
($\sigma$ = 0.18\arcdeg$\pm$0.02\arcdeg) located at
$\alpha$=10$^{\rm h}$23$^{\rm m}$18$^{\rm s}$ and $\delta$=--57\arcdeg45\arcm50\arcs\ (J2000). The morphological and spectral analyses of the
source did not allow a firm identification of the origin of the VHE
emission.  
Energy-dependent morphology studies have proven to be a powerful tool to
distinguish between alternative source models and respective
VHE $\gamma$-ray emission signatures \citep{HESSJ1825}. In order to discriminate among the different $\gamma$-ray emission
scenarios proposed and to investigate energy-dependent morphology, follow-up observations
were pursued. The region was observed for an additional 33 h in
January and March 2007, April and May 2008, and May to June
2009. The complete data set is presented here and has a  live time of 45.9~h
after quality cuts to remove data taken during bad weather
conditions or hardware irregularities \citep{HESSCrab}. The data were taken in
wobble-mode configuration, where the telescopes are
pointed offset (from 0.5\arcdeg\ to 2.5\arcdeg\ in this data set) from the
nominal source location to allow simultaneous background
estimation. The zenith angles range from 30\arcdeg\ to 55\arcdeg, with a mean
value of 38\arcdeg, which results in a mean energy threshold of
$\sim$0.8~TeV when using \emph{hard cuts} on the EAS image size of 200
photoelectrons (p.e.). The mean energy threshold of the new data set is slightly
higher than the one in the previous analysis due to lower reflectivity
of the IACT mirrors compared to the 2006 observations and due to a
stricter cut on the image size than the one used for the 2006 data set
(\emph{standard cuts} at 80 p.e.). The optical
response of the system was estimated from the Cherenkov light of
single muons hitting the telescopes as explained in \cite{HESSCrab}.

\begin{figure}[t!]
 \centering
 \includegraphics[width=0.5\textwidth]{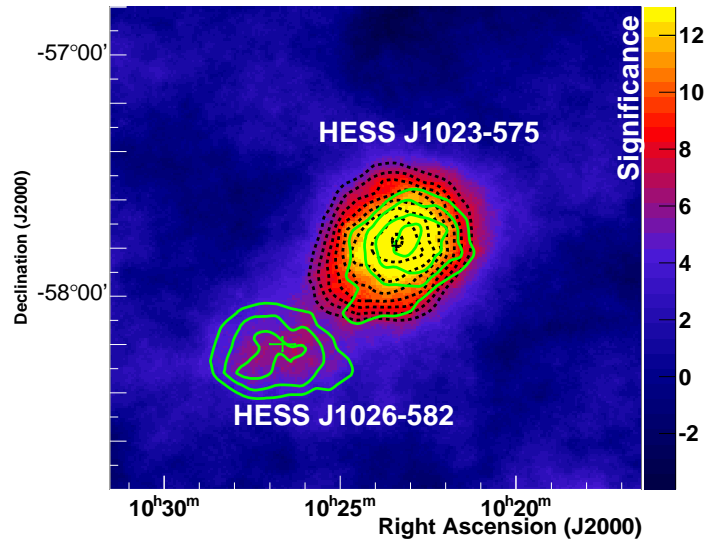}
 \caption{Image  of the Westerlund 2 field showing the significance of the VHE $\gamma$-ray
   emission using an oversampling radius of 0.22$^{\rm o}$. The contours in black (dashed) and green (solid) show the significance levels for the maps at low and high energy range, respectively, using the same oversampling radius, starting at 5$\sigma$ in steps of 1$\sigma$. The maximum of the color scale has been chosen such that the typical detection threshold $\sim$5$\sigma$ appears on the transition from blue to red. }
 \label{fig1:W2AllE}
\end{figure}

\begin{figure*}[t!]
\centering
\includegraphics[width=0.47\textwidth]{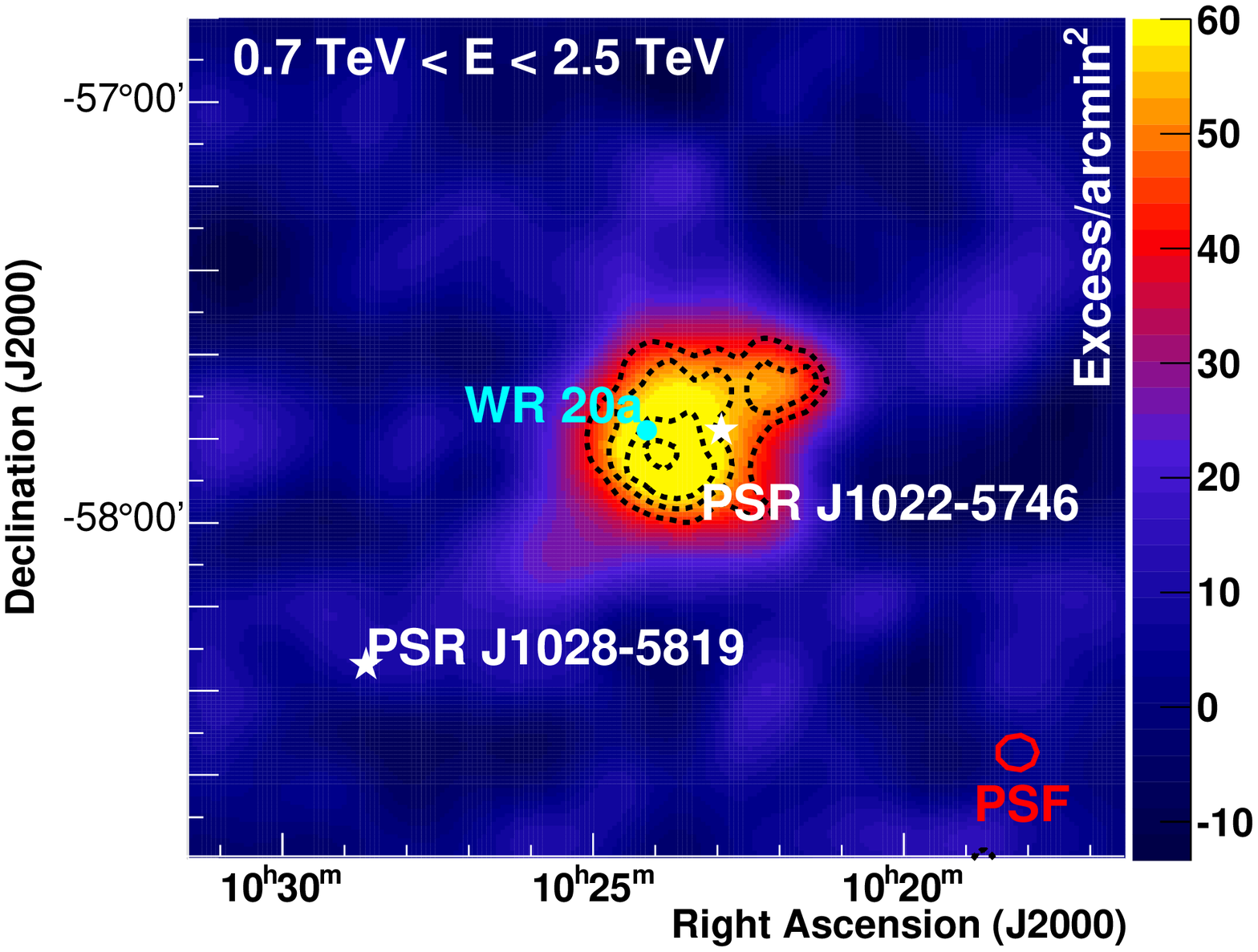}
\includegraphics[width=0.47\textwidth]{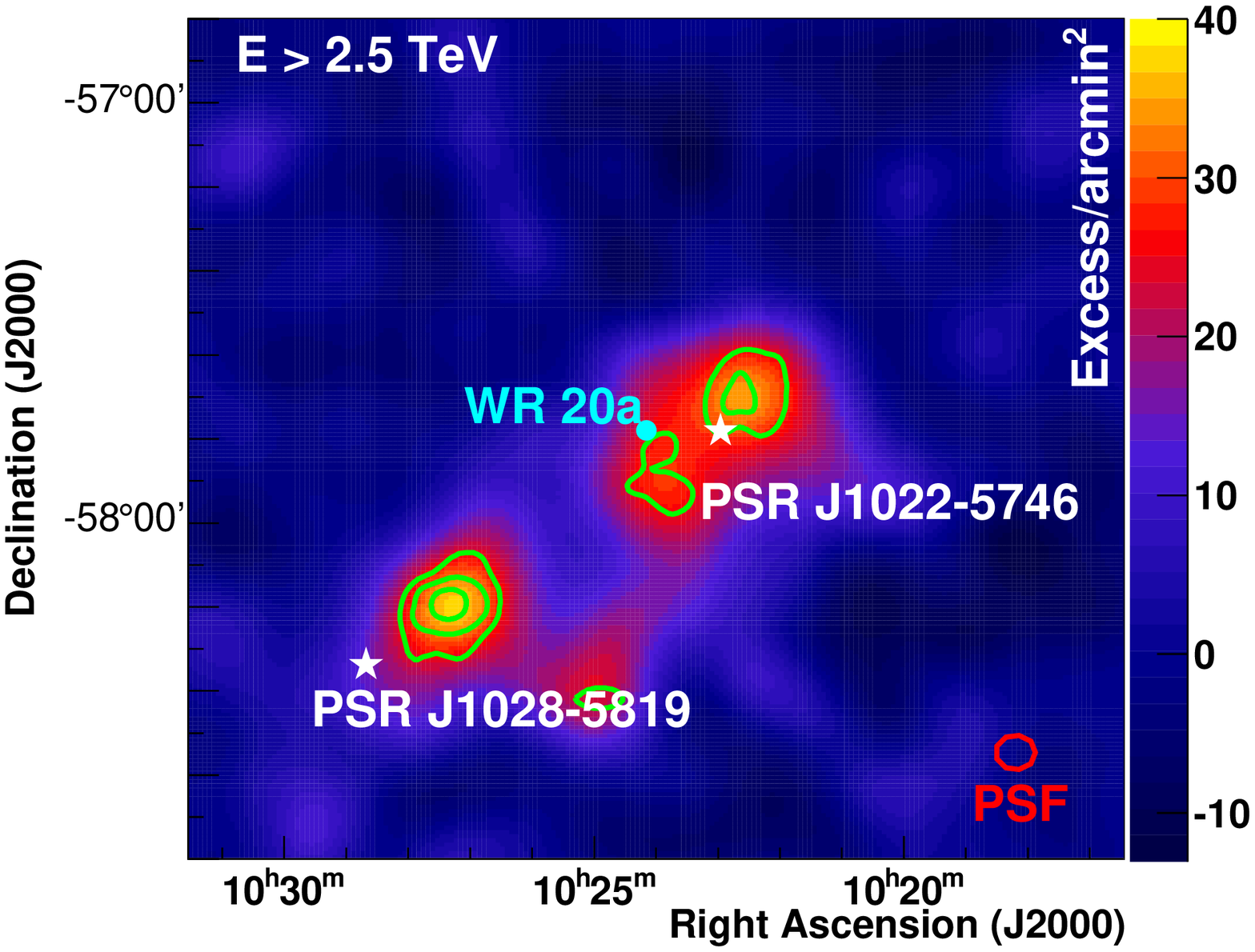}
\caption{Gaussian-smoothed ($\sigma$=0.08$^{\rm o}$) excess images
  for the Westerlund 2/RCW~49 and PSR\,J1028--5819 region. On the left,
  the low energy map (0.7~TeV~$<$E~$<$~2.5~TeV) is shown, while the
  high energy map (E~$\geq$~2.5~TeV ) is displayed on the right. The
  H.E.S.S. significance contours (dashed black lines for the low energy map and solid green ones for the high energy map) are calculated using an
  oversampling radius of 0.1$^{\rm o}$ and are shown above 4$\sigma$
  in steps of 1$\sigma$. The position of WR 20a and the two Fermi LAT pulsars are marked in cyan and white (for more
  details see text).}
\label{fig2:W2Energies}
\end{figure*}

The data have been analyzed using the H.E.S.S. standard event
reconstruction scheme \citep{HESSCrab} using the Hillas second moment
method \citep{Hillas}. A second independent analysis chain was used to cross-check the results, including independent calibration of pixel amplitudes
and identification of problematic or dead pixels in the IACTs cameras, leading to
compatible results. To achieve a maximum signal-to-noise ratio and
to improve angular resolution, \emph{hard cuts} were used to reduce the
data, and only observations runs (of 28 min duration) with four
active IACTs were used to study the morphology and spectrum of the
source. 

The large energy range covered by H.E.S.S. allows the investigation of the morphology and spectral features in
the region of interest in different energy bands. For this analysis,
two energy bands were selected \emph{a priori} such that each band has a similar
signal-to-noise ratio, for the range of
zenith angle of the current observations, assuming the source
has a photon index of 2.2 (a typical value for Galactic VHE
$\gamma$-ray sources).  Images of the VHE $\gamma$-ray excess (see
next Sect.), corrected by the corresponding radial acceptance in each
energy band, were produced for two bands (0.7$<$ E$<$2.5 TeV and E$\geq$2.5 TeV).

\subsection{Energy-Dependent Morphology}

Fig.~\ref{fig1:W2AllE} shows the significance map of the region using
an oversampling radius of 0.22$^{\rm o}$ (optimized for extended
sources, \citealt{Survey06}). The background in each bin of the image
was estimated using a ring with a mean radius of 1$^{\rm o}$ (as defined in \citealt{HESSCrab}) around the test position. The previously reported source
HESS\,J1023--575 is detected with a peak significance of 16$\sigma$ corresponding to 545 excess events above background.
A second excess appears to the south-east towards the direction of PSR\,J1028--5819 (dubbed
HESS\,J1026--582), showing a peak significance of 7$\sigma$ pre-trials (169
excess events). 
Source locations are extracted from the uncorrelated excess map fitted to a
two-dimensional Gaussian function folded with the H.E.S.S. point spread
function (PSF; $\sim$0.08$^{\rm o}$ above 200 p.e.). The best fit
position lies at $\alpha$ = 10$^{\rm h}$23$^{\rm m}$24$^{\rm s}\pm$7.2$^{\rm s}_{\rm stat}$ and $\delta$ =
--57$^{\rm o}$47\arcm24\arcs$\pm$1\arcm12\arcs$_{\rm stat}$  (J2000) for HESS\,J1023--575,
compatible with the previously reported position. For the second
source, HESS\,J1026--582, the Gaussian peaks at $\alpha$ = 10$^{\rm
  h}$26$^{\rm m}$38.4$^{\rm s}\pm$21.6$^{\rm s}_{\rm stat}$ and
$\delta$ = --58$^{\rm o}$12\arcm00\arcs$\pm$1\arcm48\arcs$_{\rm stat}$
(J2000), 0.6$^{\rm o}$ apart from the other source. The systematic error is estimated
to be 20\arcs\ per axis \citep{AngularRes}. The two derived positions and
statistical errors are marked with black and green crosses in
Fig. \ref{fig1:W2AllE}. 

To further investigate the nature of the two detected excesses, the image
has been analyzed in the two energy bands described above. For the
lower energy band, between 0.7 and 2.5 TeV, the image
(Fig. \ref{fig2:W2Energies}) shows a distinct excess emission at the
location of HESS\,J1023--575, extended beyond the nominal size of the
H.E.S.S. instrument PSF, with an intrinsic extension of
$\sigma$=0.18$^{\rm o}\pm$0.02\arcdeg. The second excess is
strongly reduced in this energy band. However, this second
emission region, HESS\,J1026--582, is clearly visible on the high
energy image (E$\geq$2.5~TeV), indicative of a very hard photon index
and also evidence for extension at the scale $\sigma$=0.14$^{\rm
  o}\pm$0.03\arcdeg. The two Gaussian-smoothed ($\sigma$=0.08\arcdeg) excess maps are
shown in Fig. \ref{fig2:W2Energies} with significance contours above
4$\sigma$ in black dashed lines for the low energy map and solid green
lines for the high energy one, obtained by using an integration
radius of 0.1\arcdeg. The significance contours for a larger integration radius
(0.22\arcdeg) for the lower (in black) and the higher (in green)
energy bands and are also shown in Fig. \ref{fig1:W2AllE} together
with the total significance map above 200 p.e.

To further investigate the interplay between the two neighboring sources as
of HESS\,J1023--575 and HESS\,J1026--582, a rectangular region (slice) is defined in the uncorrelated excess maps along the
line connecting the best-fitted positions of the two spots with a
width of twice the H.E.S.S. PSF (see Fig.~\ref{fig3:RadialProfile}
inset). The profile resulting from a projection of the slice on the long axis is shown in Fig.~\ref{fig3:RadialProfile} for the low (in red)
and high (in blue) energy images. The radial profile for the
low energy map is well fitted by a single Gaussian function
($\chi^2$/$\nu$=5.57/6, corresponding to a probability of P=0.47)
centered at a position compatible with the centroid of
HESS\,J1023--575 and the position of Westerlund~2 and PSR\,J1022--5746 (cyan and
white markers respectively in the inset figure). Their positions with respect to the center of the slice are marked with dashed
lines in the profile figure. However, the profile corresponding to the
high energy events shows a second peak towards the direction of
PSR\,J1028--5819 (dashed line) and a fit to a double Gaussian function
is clearly favored ($\chi^2$/$\nu$=0.45/4, P=0.97) against a single Gaussian one
($\chi^2$/$\nu$=10.88/6, P=0.09). 

Based on the limited statistics at E $\geq$ 2.5 TeV, the peak of the VHE
emission related to HESS\,J1023--575 appears to shift towards the
location of the LAT pulsar PSR\,J1022--5746 at higher energies
although the statistic is too scarce to resolve further
energy-dependent morphologies. The angular separation of the two
best-fitted positions for the low and high energy images is
$\sim$0.08\arcdeg, of the order of the H.E.S.S. PSF.

\subsection{Spectral Analysis}

Spectral information has been obtained in the following way: Two circular regions were
defined around the respective centroids of HESS\,J1023--575 and
HESS\,J1026--582, with radius 0.33\arcdeg. The background was evaluated by using the {\it reflected
background method}, in which symmetric regions, not
contaminated by the sources, are used to extract the background
\citep{HESSCrab}. The energy spectra are derived by means of a
forward-folding maximum likelihood fit \citep{CATSpectrum}.

The two energy spectra are shown in Fig. \ref{fig4:spectra}. 
The spectrum of the region containing HESS\,J1023--575 is well fit
by a power law function dN/dE=N${\rm_0}$(E/1~TeV)$^{-\Gamma}$, with a photon index of $\Gamma$=2.58$\pm$0.19$_{\rm~stat}\pm$0.2$_{\rm~sys}$ and a
normalization constant of N$_{\rm 0}$=(3.25$\pm$0.50$_{\rm stat}$)$\times$10$^{-12}$~TeV$^{-1}$cm$^{-2}$s$^{-1}$ (red markers and line), confirming the
results from the previous analysis (dashed line) \citep{HESSJ1023m575}. The previously
measured flux level was slightly higher but this difference is explained by the larger
integration radius (0.39\arcdeg) used previously. At high energies the previous flux of HESS\,J1023--575 was contaminated with some fraction of
the events related to the new source HESS\,J1026--582.
 The energy spectrum of HESS\,J1026--582 is also well represented by a power law function
(blue markers and line), showing a hard photon index of
$\Gamma$=1.94$\pm$0.20$_{\rm~stat}\pm$0.2$_{\rm~sys}$ as expected from
the energy maps, with a normalization constant of N$_{\rm
  0}$=(0.99$\pm$0.34$_{\rm stat}$)$\times$10$^{-12}$~TeV$^{-1}$cm$^{-2}$s$^{-1}$. The
systematic error on the normalization constant N$_{\rm 0}$ is estimated from
simulated data to be 20\% \citep{HESSCrab}.

The two measured photon indices in the energy range between 1 and 10
TeV differ by $\Delta\Gamma$=0.7, which taking into account their
statistical errors, lead to an incompatibility of 2.5$\sigma$ for the two VHE
emission regions. The different spectral indices as well as the clear
separation of the two sources in the images favor the interpretation
of the two emission regions as two independent sources rather than
having a common origin.

\section{Discussion}

Follow up observation on the previously discovered source
HESS\,J1023--575 lead to confirmation of the features of this source as well as the
discovery of a new VHE $\gamma$-ray source, HESS\,J1026--582.  The two
TeV emission regions appear spatially distinct and show different
spectral characteristics, favoring an interpretation of being two independent
sources. 
Accordingly, different scenarios regarding the nature of both sources
in the context of being related to energetic pulsars or massive stars
and their winds are discussed in the following.

\subsection{Emission scenarios related to pulsars and their winds}
The spatial coincidence with recently discovered energetic $\gamma$-ray
pulsars motivates the investigation of scenarios where $\gamma$-ray pulsars
are powering a relativistic wind nebula. 
In this scenario particles are accelerated to very high energies along
their propagation into the pulsar surroundings or at the shocks produced in
collisions of the winds with the surrounding medium. As a result of
the interactions of relativistic leptons with the magnetic field and
low energy radiation (of synchrotron origin, thermal, or microwave
background), non-thermal radiation is produced from the lowest
possible energies up to $\sim$100 TeV. On the other hand, for magnetic fields of a few $\mu$G, young electrons create a small synchrotron
nebula around the pulsar which should be visible in X-rays, in
contrast to a much larger TeV nebula, generated by the inverse Compton
(IC) process (for a recent review see \citealt{pwn}). 
Typically only pulsars with high spin-down energy ($\sim$10$^{\rm
  33}$~erg s$^{\rm -1}$, \citealt{Pavlov}) produce prominent pulsar wind
nebulae (PWNe). Both pulsars reported from Fermi LAT observations at
energies below the H.E.S.S. detection threshold show high spin-down
luminosity and features similar to other pulsars previously associated
with VHE $\gamma$-ray sources such as HESS\,J1825--137
\citep{HESSJ1825}.

\subsubsection{HESS\,J1023-575}

The LAT-discovered $\gamma$-ray pulsar PSR\,J1022--5746 is the
youngest and most energetic among the $\gamma$-ray only pulsars known
to date, being only one order of magnitude weaker than the Crab
pulsar. An X-ray source, CXOU J102302.8--574606.9 \citep{chandrapsr}, found in a 130~ksec \emph{Chandra} observation has been suggested as a potential
counterpart, 8\arcm\ apart from the Westerlund 2 core, and its relation to
the stellar cluster thus remains unknown.  A distance to CXOU J102302.8--574606.9 of $>$10 kpc was proposed based on the column density from Chandra observations \citep{Fermisymp}. The fitted
column density N$_{\rm H}\sim$1.3$\times$10$^{\rm 22}$ cm$^{\rm -3}$
indicates a large distance d$>$8~kpc, although given the uncertainties
in the measurement and the large absorption expected in the line of
sight towards Westerlund 2, a possible association with the open
cluster cannot be disregarded. 
On the other hand, no extended synchrotron PWN has been found yet
albeit \emph{Suzaku} observation time was dedicated to directly investigate
this hypothesis \citep{SUZAKU}. As noted by \cite{SUZAKU} the ratio of TeV to X-ray luminosity is larger than one, assuming that the size of the X-ray
emission they considered to obtain upper limits on the
diffuse non-thermal emission in the 0.7 to 10 keV range corresponds to
the same emitting region at VHE. That would suggest that the population of TeV
emitting particle has already cooled and does not shine significantly
in X ray synchrotrons as is the case of most middle aged TeV nebulae \citep{Okkie}.

With its high spin-down luminosity, the Fermi LAT pulsar is
sufficiently strong to power the observed VHE $\gamma$-ray
emission. Assuming the pulsar is at a distance of 8 kpc, the
luminosity of the H.E.S.S. source in the 1 to 10~TeV range is 4.8$\times$10$^{\rm
  34}$ erg s$^{\rm -1}$. 
The association with the pulsar would
imply conversion from rotational energy into non-thermal emission with
efficiency 0.4\%, comparable to those inferred for other VHE
$\gamma$-ray sources associated with PWNe, i.e., HESS\,J1912+1011
\citep{HESS1912} or HESS\,J1718--385 \citep{HESS1718}. However, the
pulsar distance and age limit the maximum extension of the nebula (see
i.e. \citealt{MHDModel}) and render it difficult to explain the total
extension of the TeV emission as due to inverse Compton scattering of
the PWN on the interstellar radiation field \citep{pwn,Okkie}. The projected size of
HESS\,J1023--575 corresponds to $\sim$28~pc(d/8.0 kpc)
(i. e. 6 or 31 pc assuming distances of 2 and 10 kpc, respectively). An
estimation of the maximum size of the nebula is given by \cite{MHDModel} (KC model):
\begin{figure}
\centering
\includegraphics[width=0.5\textwidth]{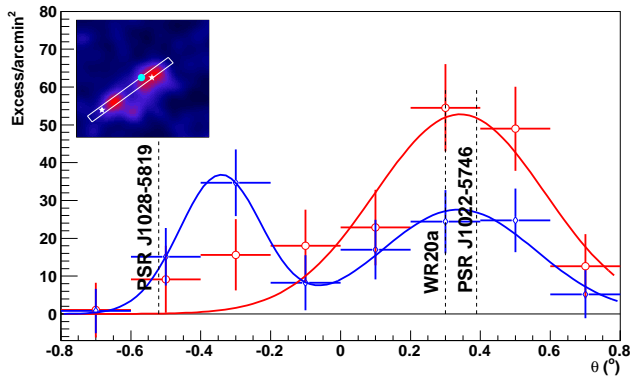}
\caption{Profile of the TeV emission along the slice defined to contain the maximum
  of the two $\gamma$-ray excesses, as illustrated in the inset. The width of
  the slice is defined to be twice the H.E.S.S. angular
  resolution. The profile on the low energy image is shown in red
  while the one for the high energy image is shown in blue. The relative positions of WR~20a and the two Fermi LAT
  pulsars with respect to the center of the slice are indicated by dashed lines.
}
\label{fig3:RadialProfile}
\end{figure}
\begin{figure}
\centering 
\includegraphics[width=0.5\textwidth]{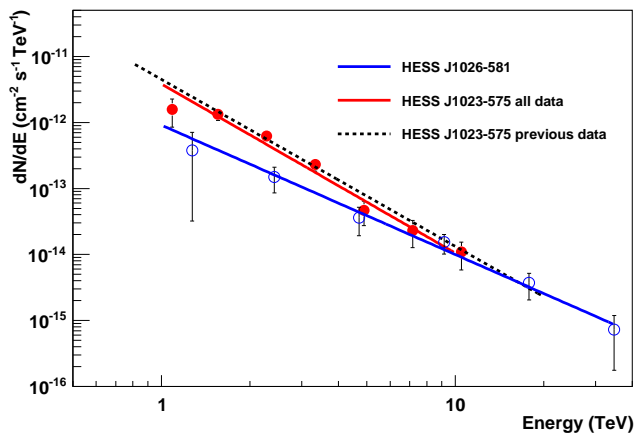}
\caption{Differential energy spectra of HESS\,J1023--575 (red filled circles) and
  dashed black line for the previous detection and HESS\,J1026--582 \citep{HESSJ1023m575}
  (blue open circles).
}
\label{fig4:spectra}
\end{figure}
\begin{eqnarray}
R\sim2\times10^{20}\left({L_{\rm SD}\over 10^{37}\rm erg/s}\right)^{1/3}\left({t\over 10^{5}\rm yr}\right)^{1/3}\nonumber \\
\left({p_{\rm ism}\over10^{-12}\rm erg/cm^3}\right)^{-1/3}{\rm cm} \sim 19~\rm pc
\label{pwn_size}
\end{eqnarray}
\begin{table*}[!t]
\caption{Properties of high energy pulsars associated to the TeV
  sources and inferred TeV luminosity for the nominal pulsar distance}              
\label{table1}      
\centering                                      
\begin{tabular}{c c c c c c c c}          
\hline\hline                        
HESS source & Pulsar Name & $\dot{E}$ [erg~s$^{\rm -1}]$ & d$_{\rm
  nom}$ [kpc] & $\tau$ [kyrs] & B [10$^{\rm 12}$ G] &
L$_{\gamma}$(d/d$_{\rm nom}$)$^{\rm 2}$(1--10~TeV) [10$^{\rm 33}$erg s$^{\rm -1}$]\\    
\hline                                   
    HESS\,J1023--575 & PSR\,J1022--5746 & 1.1$\times$10$^{\rm 37}$ & 8 & 4.6 & 6.6 & 48 \\
   HESS\,J1026--582 & PSR\,J1028--5819 & 8.3$\times$10$^{\rm 35}$ & 2.3
   & 89  &  1.2 & 2.5  \\     
\hline
\end{tabular}
\end{table*}
obtained by balancing the total luminosity at present provided by the
pulsar against the interstellar medium pressure. The obtained value  is
1.5 times smaller than the estimated size of the H.E.S.S. source for a
pulsar as far as 8 kpc (28 pc), although this value has a strong dependency on
the assumed input parameters. Under these assumptions (using the
pulsar parameters and a typical pressure for the ISM of $p_{\rm
  ism}\sim$2$\times$10$^{\rm -12}$ eV~cm$^{\rm -3}$), the KC PWN
model for a pulsar located at the nominal distance of Westerlund~2 does not describe well the VHE emission and more
sophisticated scenario are needed, including possible anisotropy of
the pulsar wind or evolution of the spin-down luminosity which
will allow a higher intrinsic spin-down energy. Also, the presence
of stellar winds from O type stars are known to create cavities,
allowing a larger PWN size compared to expansion into more typical ISM
pressures. A PWN according to the model of KC would be consistent
with a distance estimate of $\sim$2 kpc. Recently, a revised distance was
estimated by \cite{FermiBlind} suggesting PSR\,J1022--5746 at a
distance of 2.4 kpc.

\subsubsection{HESS\,J1026--582}

The association of the VHE $\gamma$-ray source
HESS\,J1026--582 with the Fermi LAT pulsar PSR\,J1028--5819
($\alpha$=10$^{\rm h}$28$^{\rm m}$39.8$^{\rm s}$ and
$\delta$=--58\arcdeg17\arcm13\arcs, J2000, with a 95\% confidence
level radius of 0.079\arcdeg) seems
likely and can be naturally described by interaction of the pulsar
wind with the surrounding medium. Moreover, an alternative counterpart
has not been proposed so far. The spin-down energy of the pulsar is
high enough to power the  VHE $\gamma$-ray emission, which, assuming a
distance of 2.3 kpc, results in a total luminosity in the 1 to 10 TeV
energy range of
2.5$\times$10$^{\rm 33}$  erg s$^{\rm -1}$. The implied conversion
efficiency from spin-down power into particle luminosity would imply a
similar value to the one considered above, about 0.3\%. No PWN
association at X-ray energies has been found so far, although observations with both
XMM-\emph{Newton} and \emph{Suzaku} have been carried out triggered by the
Fermi LAT detection. The corresponding projected extension of
HESS\,J1026--582 is 5.6 pc  for an assumed distance of 2.3 kpc. Using
Eq. \ref{pwn_size}, the estimated maximum size is $\sim$8~pc, which is
compatible with the observed extension at VHE energies. The large
offset of the pulsar with respect to the center of gravity of the VHE
source (0.28\arcdeg) and asymmetric morphology is similar to other
relic TeV PWN such as HESS\,J1825--137 and can be explained either by
the proper velocity of the pulsar or by an asymmetric reverse shock
resulting from a supernova that exploded initially in an inhomogeneous
medium \citep{offset}. If the pulsar was born at the H.E.S.S. source
position, a pulsar transverse velocity of  $\sim$100 km s$^{-1}$ can
be estimated using the dispersion-measured based distance of 2.3
kpc, and a characteristic age of 8.9$\times$10$^{4}$ yrs. This is in
the observed range of known pulsar velocities. 

A summary of the observed parameters of the high energy pulsars and
inferred luminosity above 1 TeV for the nominal distances (d$_{\rm
  nom}$) of the associated TeV sources is given in Table \ref{table1}.

\subsection{Emission scenarios related to massive stars and their winds}

\begin{figure}
\centering
\includegraphics[width=0.4\textwidth,height=70mm]{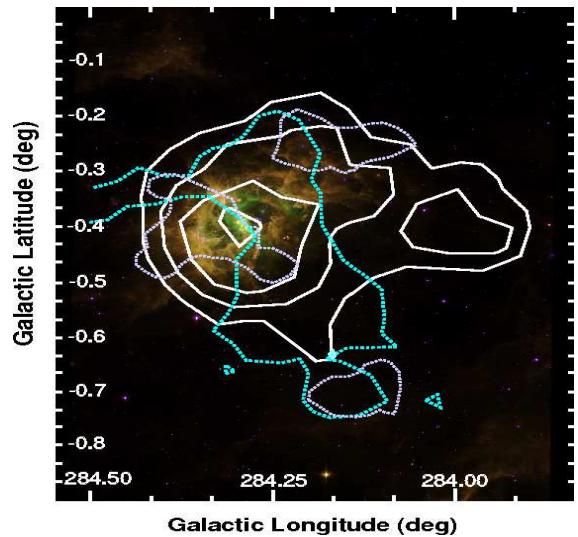}
\caption{Spitzer/IRAC GLIMPSE three-color image (in red the 5.8 $\mu$m image, 8 $\mu$m in green and 3.6$\mu$m in blue) of RCW 49 overlaid with $^{\rm
    12}$CO(J~=~2--1) contours at velocity ranges of 10.8 to 20.9 km~s$^{\rm -1}$ and 1.2 to 8.7 km~s$^{\rm -1}$ observed with the Nanten 2 telescope (in blue and purple respectively). The white contours show the H.E.S.S. significance derived for a correlation radius of 0.1\arcdeg, hard cuts and a maximum energy of 2.5~TeV. The lowest contour corresponds to 4~$\sigma$.}
\label{fig5:MC}
\end{figure}

The possibility of an association of HESS\,J1023--575 with the open
cluster Westerlund~2 was discussed in the HESS\,J1023--575 discovery
paper \citep{HESSJ1023m575}. 
Different mechanisms involving cosmic-rays accelerated in expanding
stellar winds or supernova blast waves interacting with the boundaries
of the interstellar radiation field or the boundaries of the radio blister \citep{Voelk1983} are among the possible scenarios to explain
the VHE $\gamma$-ray emission. The total mechanical luminosity (W$_{\rm total}$)
available from stellar winds in Westerlund 2 is estimated to be 3.6
$\times$ 10$^{51}$ erg \citep{Rauw07}. Observations of the molecular cloud content in the region by \cite{DameCO} shows a complex and rich
environment and triggered high resolution (1.5\arcm) observations with the
Nanten 2 4-m sub-mm telescope. The latter observations resolved
massive molecular clouds at $\sim$16~km~s$^{-1}$ and
$\sim$4~km~s$^{-1}$ (6.5 and 5.2 kpc), that were previously reported by \cite{DameCO}, as
well as a second component at $\sim$--4~km~s$^{-1}$ which appears to be
coincident with a region of ongoing star formation. To illustrate the
extent of the star formation region, a three-color Spitzer/IRAC
GLIMPSE image is shown in Fig. \ref{fig5:MC}, in red the 5.8 $\mu$m
image, 8 $\mu$m in green and 3.6 $\mu$m in blue. The Nanten 2 radio signal reported in the velocity ranges of 10.8 to 20.9 km~s$^{\rm -1}$ and 1.2 to 8.7 km~s$^{\rm -1}$ is overlaid in blue and purple dashed lines respectively, showing a strong spatial correlation with the cluster. The H.E.S.S. low energy significance contours are shown in white. These GMCs 
provide a sufficiently dense target ($\sim$8--9$\times$10$^{4}$M$_{\odot}$ which translates to a density of n$\sim$54--60~cm$^{-3}$ for a size corresponding to the HESS\,J1023-575 extension at 8 kpc) for high-energy
particle interactions, allowing the production of $\gamma$-rays from $\pi$-decay
 via inelastic pp-collisions. In both cases, considering
the GMC mass and distance, (and using W$_{\rm pp}$=L$_{\gamma}~\tau_{\rm pp}$(n) , with $\tau_{\rm pp}$ being the cooling time for p-p interactions) 
less than 0.05$\%$ of the total mechanical luminosity estimated for the stellar winds in Westerlund~2
would be sufficient to explain the observed VHE emission. 

A power-law extrapolation of the H.E.S.S. spectrum to lower energies (to 30 GeV) yields a flux of 2.8$\times$10$^{-8}$ TeV$^{-1}$ cm$^{-2}$ s$^{-1
}$ (or 1.7--6.2$\times$10$^{-8}$  TeV$^{-1}$ cm$^{-2}$ s$^{-1 }$
considering statistical errors in the normalization and spectral index). 
This flux extrapolation is well above the differential flux measured
by Fermi LAT at approximately the same energy ($\sim$3.3$\times$10$^{-9}$
TeV$^{-1}$ cm$^{-2}$ s$^{-1 }$, \citealt{fermi1cat}). Thus, the combination of the
H.E.S.S. and Fermi LAT results implies a spectral hardening towards
lower energies, which can be described with a turnover of the
H.E.S.S. spectrum at $\sim$100 GeV. If hadronic emission processes are responsible for the
emission, the lower energy cutoff could result from cosmic-rays of
energy $<$1~TeV being prevented from reaching the GMC, perhaps via
slow diffusion as demonstrated by \cite{AharonianAtoyan96}, where only
high energy hadrons reach the surrounding molecular clouds. In this
scenario, considering an age for Westerlund~2 of 1 to 2 million years, low diffusion
coefficients in the order of 10$^{26}$~cm$^2$ s$^{-1}$ are necessary to
prevent low energy particles from reaching the $\gamma$-ray production
site.

Emission scenarios where the bulk of the VHE $\gamma$-rays is produced
close to the massive WR stars WR\,20a and/or WR\,20b were
already disfavored given the observed extension and lack of
variability in the VHE $\gamma$-ray emission, as discussed in
\cite{HESSJ1023m575}. However, $\gamma$-rays from collective stellar
wind interactions, where energetic particles experience multiple
shocks, may still contribute to the observed VHE emission.

\section{Conclusions}

The H.E.S.S. source HESS\,J1023--575 has been re-observed in order to
search for energy-dependent morphology, that could provide clues on
the origin of the  $\gamma$-ray emission. 
The analysis of the new data unveils a second VHE emission peak
  towards the direction of the energetic high energy pulsar
  PSR\,J1028--5819, reported recently from Fermi LAT observations at lower energies (0.1 to $\sim$30 GeV).
The centers of gravity of the two VHE $\gamma$-ray emission regions are
well separated (0.6\arcdeg) and the radial profile shows two clear
peaks at the higher energies, while at lower energies the VHE emission
corresponding to HESS\,J1026--582 appears very much suppressed. This
new source is detected with a significance level of 7$\sigma$ and is
characterized by a very hard photon spectral index of 1.9.
HESS\,J1026--582 is preferably interpreted as due to a PWN energized by PSR\,J1028--5819, with
similar characteristics as those previously detected by H.E.S.S. (such
as hard spectrum photon index, offset of the high energy pulsar with
respect the center of the TeV emission, similar conversion
efficiency from rotational energy into non-thermal emission,
etc). Further observations with X-ray satellites of this region will
have a chance to provide the required additional information to
support and settle the suggested scenario of being another GeV Pulsar/TeV
PWN association. 

The previously reported VHE source HESS\,J1023--575 has been confirmed 
and the analysis of the new data shows compatible
results. Multi-wavelength observations of the regions provided new
insights on the potential origin of the $\gamma$-ray emission as
  discussed in the previous section. A possible association with the
recently discovered $\gamma$-ray PSR\,J1022--5746
(1FGL\,J1023.0--5746), is supported by spatial coincidence with
HESS\,J1023--575, and can be easily accommodated in terms of
energetics if the pulsar is as close as 2--3 kpc, while more
sophisticated PWN models are necessary if it is located significantly further
away. On the other hand, the molecular content of the region provides
sufficient target material to explain the emission through hadronic
interaction of cosmic-rays accelerated by a range of feasible
mechanisms in the open cluster interacting with molecular clouds. 

Whether the emission is due to the not-yet-detected PWN associated to
the Fermi LAT pulsar PSR\,J1022--5746 or mechanisms related to
acceleration of cosmic-rays in the open cluster, a spectral flattening
at lower energies must be present to explain the lower MeV-GeV
flux level. 
In the first case, IC processes would naturally reproduce the spectral
shape, while for hadronic emission, a lower cutoff at a several tens of GeV
would reflect a lower cutoff on the cosmic-ray spectrum at $\sim$ 1
TeV. The latter scenario could be explained by assuming a low
diffusion coefficient in the region, which would not be surprising in
the context of the cavities created by the stellar winds from the
young OB stars in the stellar cluster. Despite the already
accomplished multi-wavelength coverage, a firm identification is still
pending, whereas the high-power pulsar PSR\,J1022--5746 is the prime counterpart hypothesis to adopt
now. Unfortunately the statistics at VHE is too scarce to
distinguish any further hint of energy-dependent morphology which
would have the potential to settle the association with the pulsar
or the open cluster. A clear identification of a PWN associated to the
Fermi LAT pulsar PSR\,J1022--5746 in X-rays would help to elucidate the
real nature of HESS\,J1023--575, but given the stellar density and
richness in X-ray sources towards Westerlund~2, this appears to be a serious
observational challenge. In particular deeper VHE observations might allow higher energy resolution studies and a firm identification of the origin of the VHE emission.

\begin{acknowledgements}
The support of the Namibian authorities and of the University of Namibia
in facilitating the construction and operation of H.E.S.S. is gratefully
acknowledged, as is the support by the German Ministry for Education and
Research (BMBF), the Max Planck Society, the French Ministry for Research,
the CNRS-IN2P3 and the Astroparticle Interdisciplinary Programme of the
CNRS, the U.K. Science and Technology Facilities Council (STFC),
the IPNP of the Charles University, the Polish Ministry of Science and 
Higher Education, the South African Department of
Science and Technology and National Research Foundation, and by the
University of Namibia. We appreciate the excellent work of the technical
support staff in Berlin, Durham, Hamburg, Heidelberg, Palaiseau, Paris,
Saclay, and in Namibia in the construction and operation of the
equipment.
\end{acknowledgements}



\end{document}